\documentclass[aps,pra,amssymb,amsmath,eqsecnum,nofootinbib]{revtex4}
\usepackage{graphicx}

\newtheorem{definition}{Definition}[section]

\newtheorem{remark}{Remark}[section]

\begin{document}
\title{The natural definition of the quantum dynamical entropy in the framework of deformation quantization}
\author{Gavriel Segre}
\begin{abstract}
It is shown how, in the framework of deformation quantization, the quantum dynamical entropy may be simply defined as the Kolmogorov-Sinai entropy of the quantum flow.
\end{abstract}
\maketitle
\newpage
\tableofcontents
\newpage
\section{Acknowledgments}
I would like to thank Gianni Jona-Lasinio and Vittorio de Alfaro for many precious teachings. Of course nobody among them has any responsibility as to any eventual mistake contained in this paper.
\newpage
\section{Deformation quantization} \label{sec:Deformation quantization}
The core of deformation quantization \cite{Sternheimer-00},
\cite{Gutt-00}, \cite{Dito-Sternheimer-02},
\cite{Schlichenmaier-05}, \cite{Sternheimer-05},
\cite{Zachos-Fairle-Curtright-05} consists in the idea that
quantization may be understood as a deformation of the structure
of the algebra of classical observables, rather than a radical
change in the nature of the observables.

To formalize the mathematical essence of its structure let us
start from the following \cite{Marsden-Ratiu-99}, \cite{Arnold-Givental-01}:
\begin{definition}
\end{definition}
\emph{symplectic manifold:}

a couple $ ( M ,  \omega )  $ where:
\begin{itemize}
    \item M is a differentiable manifold
    \item $ \omega $ is a symplectic  (i.e.  closed,
    non-degenerate) 2-form over M
\end{itemize}

Given a symplectic manifold $ ( M , \omega ) $ and a map $ f \in C^{\infty} (M) $:
\begin{definition}
\end{definition}
\emph{hamiltonian vector field generated by f:}
\begin{equation}
    X_{f} \in \Gamma ( T M )  \, : \, i_{X_{f}} \omega = d f
\end{equation}

Given two maps $ f,
g \in C^{\infty} (M) $:
\begin{definition} \label{def:Poisson brackets}
\end{definition}
\emph{Poisson bracket of f and g:}
\begin{equation}
    \{ f , g \} \; = \; \omega ( X_{f} , X_{g} )
\end{equation}

Let us recall that  $  ( C^{\infty} (M) , \{ \cdot , \cdot \} ) $
is a Lie algebra.

Let us then define the following:
\begin{definition}
\end{definition}
\emph{$ \star$-product of  f and g: }
\begin{equation}
      f  \star g  \; := \;  \exp [ \frac{i \hbar}{2} \{ f , g\} ]
\end{equation}

The $ \star$-product may be used to introduce the following:
\begin{definition}
\end{definition}
\emph{Moyal bracket of f and g:}
\begin{equation}
 \{ f , g \}_{\hbar} \; := \; \frac{f \star g - g  \star f}{i \hbar}
\end{equation}

$ ( C^{\infty} (M) , \{ \cdot , \cdot \}_{\hbar} )
$ is called the \emph{deformation quantization} of the symplectic
manifold $ ( M , \omega ) $.

Let us then define the following:
\begin{definition}
\end{definition}
\emph{quantum flow \footnote{Except from the cases in which the Schr\"{o}dinger picture will be explicitly indicated we will work in the Heisenberg picture in which observables evolve with time.} generated by $ H \in C^{\infty}(M)$:}

the flow $ \{  U_{t}(H) \}_{t \in \mathbb{R}} $ over $ C^{\infty}(M) $ ruled by the following \emph{Moyal equation:}
\begin{equation}
    \frac{ \partial f }{ \partial t } \; = \;  \{ H , f \}_{\hbar}
\end{equation}

The classical limit consists in the fact that, for $ \hbar
\rightarrow 0 $, the Moyal bracket reduces to the Poisson bracket
and hence the Moyal equation reduces to the \emph{Liouville equation:}
\begin{equation}
    \frac{ \partial f }{ \partial t } \; = \;  \{ H , f \}
\end{equation}

\newpage
\section{Kolmogorov-Sinai entropy of a flow}
Let $ ( X , \sigma , \mu ) $ be a classical probability space and
let us introduce the following:
\begin{definition} \label{def:partition of a classical probability space}
\end{definition}
\emph{finite partitions of  $ ( X , \sigma , \mu ) $:}
\begin{multline}
    {\mathcal{P}} ( X , \sigma , \mu ) \; := \; \{ P = \{ A_{i} \}_{i=1}^{n(P)} \, : n(P) \in  {\mathbb{N}} \; , \;  A_{i} \in \sigma  \, i=1 , \cdots, n(P) \; \\
      A_{i}  \, \cap \, A_{j} \, = \, \emptyset \; i,j =1 , \cdots, n(P) \, : i \neq j  \; \mu (X - \cup_{i=1}^{n(P)} A_{i}) = 0  \}
\end{multline}

\smallskip

\begin{remark}
\end{remark}

Beside its abstract, mathematical formalization, the definition
\ref{def:partition of a classical probability space} has a precise
operational meaning.

Given the classical probability space $  ( X , \sigma ,  \mu ) $
let us suppose to make an experiment on the probabilistic universe
it describes using an instrument whose resolutive power is limited
in that it is not able to distinguish events belonging to the same
atom of a partition $ P = \{ A_{i} \}_{i=1}^{n} \in \mathcal{P} (
X , \mu ) $.

Consequentially the outcome of such an experiment will be a number
\begin{equation}
  r \in \{ 1 , \cdots , n \}
\end{equation}
specifying the observed atom $ A_{r} $ in our coarse-grained
observation of $ ( X , \sigma , \mu ) $.

We will call such an experiment an \emph{operational observation
of $ ( X , \sigma , \mu ) $ through the finite partition P} or,
more concisely, a \emph{P-experiment}.

\smallskip

The probabilistic structure of the operational observation of $ (
X , \sigma , \mu ) $ through a finite partition   $ P \in
{\mathcal{P}} ( X , \sigma , \mu ) $ is enclosed in the following:
\begin{definition}
\end{definition}
\emph{probability measure of the P-experiment}:
\begin{equation*}
    \mu_{P} \; := \; \mu |_{\sigma (P)}
\end{equation*}
where $ \sigma (P) \subset  \sigma $ is the $ \sigma$-algebra
generated by P.

\smallskip

\begin{definition}
\end{definition}
\emph{coarsest refinement of $ A = \{ A_{i} \}_{i=1}^{n} $ and  $
B = \{ B_{j} \}_{j=1}^{m} \in {\mathcal{P}}(  X , \sigma ,  \mu )
$}:
\begin{equation}
  \begin{split}
      A \, & \vee \, B \; \in {\mathcal{P}}( X  , \sigma ,  \mu )  \\
      A \, & \vee \, B \; \equiv \; \{ \, A_{i} \, \cap \, B_{j} \, \;  i =1 , \cdots , n \; j = 1 , \cdots , m  \}
  \end{split}
\end{equation}

Clearly $ \mathcal{P} ( X  , \sigma ,   \mu ) $ is closed under
coarsest refinements.

\smallskip

Let us now introduce the following:
\begin{definition}
\end{definition}
\emph{entropy of $ P = \{ A_{i} \}_{i=1}^{n(p)} \in {\mathcal{P}}
( X , \sigma , \mu ) $}:
\begin{equation}
    H(P) \; := \; - \sum_{i=1}^{n(P)} \mu_{P}( A_{i} ) \log_{2} \mu_{P}( A_{i} )
\end{equation}

\smallskip

The more abstract definition of a classical dynamical system is
the following:
\newpage
\begin{definition}
\end{definition}
\emph{classical dynamical system} :

a couple $ ( ( X , \sigma , \mu) , T) $ such that:
\begin{itemize}
    \item $ ( X , \sigma , \mu)  $ is a classical probability
    space
    \item $ T : X \mapsto X $ is such that:
\begin{equation}
    \mu \circ T^{- 1} \; = \; \mu
\end{equation}
\end{itemize}

Given  a classical dynamical system $  CDS \, := \, ( (X \, ,
\sigma , \mu)  ,  T ) $, the $T^{-1}$-invariance of $ \mu $
implies that the  finite partitions $ P = \{ A_{i} \}_{i=1}^{n} \in
\mathcal{P} ( X , \sigma , \mu ) $ and $ T^{-1}P $ have equal
probabilistic structure. Consequentially the \emph{P experiment}
and the \emph{ $T^{-1}P $-experiment} are replicas, not
necessarily independent, of the same experiment made at
successive times.

In the same way the \emph{$ \vee_{k=0}^{n-1} \, T^{-k} P
$-experiment} is the compound experiment consisting in n
replications $ P \, , \, T^{-1} P \, , \, , \cdots , \,
T^{-(n-1)}P $ of the experiment corresponding to $ A \in
{\mathcal{P}}(X , \sigma , \mu) $.

The rate of classical information for replication we obtain in
this compound experiment is clearly:
\begin{equation*}
  \frac{1}{n} \, H(\vee_{k=0}^{n-1} \, T^{-k} P )
\end{equation*}
It may be proved (cfr. e.g. the second paragraph of the third
chapter of \cite{Kornfeld-Sinai-Vershik-00}) that when n grows
this rate of classical information acquired for replication
converges, so that the following quantity:
\begin{equation}
  h( P , T ) \; := \; lim_{n \rightarrow \infty} \,  \frac{1}{n} \, H(\vee_{k=0}^{n-1} \, T^{-k} P )
\end{equation}
does exist.

Clearly $ h( P , T ) $ gives the
asymptotic rate of production of classical information for
replication of the P-experiment.

\begin{definition} \label{def:Kolmogorov-Sinai's entropy}
\end{definition}
\emph{Kolmogorov-Sinai entropy of CDS:}
\begin{equation}
  h(CDS) \; := \; sup_{P \in {\mathcal{P}}(X , \sigma , \mu)} \, h( P , T )
\end{equation}
By definition we have clearly that:
\begin{equation}
  h(CDS) \; \geq \; 0
\end{equation}
\begin{definition} \label{def:classical chaoticity}
\end{definition}
\emph{CDS  is chaotic}:
\begin{equation}
  h(CDS) \; > \; 0
\end{equation}

Let us now introduce the straightforward generalization of these
notions to flows:
\begin{definition}
\end{definition}
\emph{classical flow:}

a couple $ ( ( X , \sigma , \mu) , \{  T_{t} \}_{t \in \mathbb{R}})$ such that:
\begin{itemize}
    \item $ (  ( X , \sigma , \mu) , T_{t} ) $ is a classical
    dynamical system for every $ t \in \mathbb{R} $
    \item $  \{ T_{t} \}_{t \in \mathbb{R}} $ is an abelian group
\end{itemize}

Given a classical flow $ F = ( ( X , \sigma , \mu) , \{  T_{t} \}_{t \in \mathbb{R}}) $ it is natural
to define its dynamical entropy in the following way :
\begin{definition} \label{def:Kolmogorov-Sinai's entropy of a flow}
\end{definition}
\emph{Kolmogorov-Sinai entropy of F:}
\begin{equation}
  h(F) \; := \; h [ ( ( X , \sigma , \mu) , T_{1} )]
\end{equation}
\newpage
\section{Quantum dynamical entropy}

The problem of characterizing correctly the notion of quantum
dynamical entropy, i.e. the quantum analogue of the
Kolmogorov-Sinai entropy, has led to a plethora of candidate
notions, the more famous ones being the
\emph{Connes-Narnhofer-Thirring entropy}
\cite{Connes-Narnhofer-Thirring-87}, \cite{Benatti-93} and the
\emph{Alicki-Fannes-Lindblad entropy} \cite{Alicki-Fannes-01}, \cite{Benatti-09}.

Let us remark, anyway, that no one of these proposals has a clear information-theoretic meaning comparable to that of the Kolmogorov-Sinai entropy.

From the viewpoint of deformation quantization, from the other
side, it appears natural to define the quantum dynamical entropy
simply as the Kolmogorov-Sinai entropy of the quantum flow  $U_{t}(H)$.

In order to implement technically such an idea we have to introduce some notion concerning algebraic dynamical systems.

Let us recall first of all the following:
\begin{definition}
\end{definition}
\emph{algebraic probability space:}

a couple $ ( A, \omega ) $ such that:
\begin{itemize}
  \item A is a $ W^{\star}$-algebra
  \item $ \omega \in S(A) $ is a state over A
\end{itemize}

The algebraic probability space $ ( A , \omega ) $ is said to be commutative whether A is commutative.

\begin{definition}
\end{definition}
\emph{algebraic dynamical system:}

a triple $ ( A , \omega , \tau ) $ such that:
\begin{itemize}
  \item $ ( A, \omega ) $  is an algebraic probability space
  \item $ \tau  $ is an endomorphism of A $ \omega$-preserving  (i.e. such that $ \omega\circ \tau = \omega $).
\end{itemize}

The algebraic dynamical system  $ ( A , \omega , \tau  ) $ is said to be commutative whether $ ( A , \omega ) $ is commutative.

A classical dynamical system $ ( ( X , \sigma , \mu ) , T ) $ may be seen as the commutative algebraic dynamical system $ ( L^{\infty} ( X , \mu ) , \omega_{\mu} , \Theta_{\tau} ) $  where:
\begin{definition}
\end{definition}
\emph{state over $ L^{\infty} ( M , \mu ) $ associated to $ \mu $:}
\begin{equation}
    \omega_{\mu} ( f) \; := \; \int_{M} f d \mu
\end{equation}
\begin{definition}
\end{definition}
\emph{endomorphism of $ L^{\infty} ( M , \mu ) $ associated to $ \tau$:}
\begin{equation}
    \Theta_{\tau} (f) \; := \; f \circ \tau^{-1}
\end{equation}

Given a commutative $ W^{\star}$-algebra A let us denote by $ \mathcal{F}(A) $ the collection of finite dimensional subalgebras of A.

Given a state $ \omega \in S(A) $ and a subalgebra $ N \in \mathcal{F}(A) $ having $  \{ n_{i} \}_{i=1}^{k}$ as minimal projections:
\begin{definition}
\end{definition}
\emph{entropy of $ \omega $ with respect to N:}
\begin{equation}
    H_{\omega}(N) \; := \; - \sum_{i=1}^{k} \omega( n_{i}) \log_{2} \omega ( n_{i} )
\end{equation}

Given $ N_{1}, N_{2} \in  \mathcal{F}(A) $:
\newpage
\begin{definition}
\end{definition}
\emph{coarsest refinement of  $ N_{1} $ and $ N_{2} $:}
\begin{center}
    the subalgebra $ N_{1} \vee N_{2} \in \mathcal{F}(A) $ having as miminal projections the product of the minimal projections of, respectively, $ N_{1} $ and $ N_{2}$
\end{center}

Given a commutative algebraic dynamical system $ ADS := ( A , \omega , \tau ) $:
\begin{definition} \label{def:algebraic Kolmogorov-Sinai entropy}
\end{definition}
\emph{Kolmogorov-Sinai entropy of ADS:}
\begin{equation}
    h(ADS) \; := \; \sup_{N \in  \mathcal{F}(A) } \lim_{n \rightarrow \infty} \frac{1}{n} H( N \vee \tau (N) \vee \cdots \vee \tau^{n-1}(N) )
\end{equation}

Let us now introduce the straightforward generalization of these notions to commutative algebraic flows.

\begin{definition}
\end{definition}
\emph{algebraic flow:}

a triple $ ( A , \omega , \{ \tau_{t}\}_{t \in \mathbb{R}} ) $ such that:
\begin{itemize}
    \item  $ ( A , \omega , \tau_{t} ) $  is an algebraic
    dynamical system for every $ t \in \mathbb{R} $
    \item $  \{ \tau_{t} \}_{t \in \mathbb{R}} $ is an abelian group
\end{itemize}

Clearly an algebraic flow is said to be commutative whether the involved $ W^{\star}$-algebra is commutative.

Given a commutative algebraic flow $ AF = ( A , \omega , \{ \tau_{t}\}_{t \in \mathbb{R}} ) $:
\begin{definition} \label{def:Kolmogorov-Sinai's entropy of an algebraic flow}
\end{definition}
\emph{Kolmogorov-Sinai entropy of AF:}
\begin{equation}
    h( AF ) \; := \; h [ ( A , \omega , \tau_{1})]
\end{equation}

\bigskip

\begin{remark}
\end{remark}
Let us remark that, given a classical flow   $ CF := ( ( X , \sigma , \mu ) , \{ \tau_{t}\}_{t \in \mathbb{R}} ) $ and the associated commutative  algebraic flow   $  AF := ( L^{\infty} ( X , \mu ) , \omega_{\mu} ,  \{ \Theta_{\tau_{t}} \}_{t \in \mathbb{R}} ) $ one has that:
\begin{equation}
    h( AF) \; = \; h (CF)
\end{equation}
Since from the other side, owing to the Gelfand isomorphism, given a commutative algebraic flow $ ( A , \omega , \{ \tau \}_{t \in \mathbb{R}} ) $ there exists a classical flow $ ( ( X , \sigma , \mu ) , \{ T_{t} \}_{t \in \mathbb{R}} ) $ such that $ A = L^{\infty} ( X , \mu )$, $ \omega = \omega_{\mu} $ and $ \tau_{t} = \Theta_{T_{t}} $, the definition  \ref{def:Kolmogorov-Sinai's entropy of an algebraic flow} is conceptually equivalent to the definition \ref{def:Kolmogorov-Sinai's entropy of a flow} and has, in particular, the clear information-theoretic meaning explained in the previous section.

\bigskip

Let us now  observe that the classical flow of an hamiltonian dynamical system having as phase space the symplectic manifold $ ( M ,
    \omega) $ and as hamiltonian $ H \in C^{\infty} (M) $  may be seen as the commutative algebraic flow $  F(H) := ( L^{\infty} ( M , \mu_{Liouville}) , \omega_{\mu_{Liouville}}
    , \lim_{\hbar \rightarrow 0 }  \{ U_{t}(H) \}_{t \in \mathbb{R}}  ) $, where $ \mu_{Liouville}$ is the following:
\begin{definition}
\end{definition}
\emph{Liouville measure on $ ( M, \omega )$:}
\begin{equation}
    \mu_{Liouville} \; := \; \wedge_{i=1}^{\frac{dim M}{2}} \omega
\end{equation}
and where the fact that the Liouville state $ \omega_{\mu_{Liouville}} $ is invariant under the hamiltonian flow $  \lim_{\hbar \rightarrow 0 }   \{ U_{t}(H) \}_{t \in \mathbb{R}} $ is owed to the fact that introduced  the following:
\newpage
\begin{definition}
\end{definition}
\emph{group of the symplectic diffeomorphisms of $ ( M , \omega)$}:
\begin{equation}
  Symp (M , \omega ) \; := \; \{ \phi \in Diff(M) \, : \,  \{  f ,  g \} \circ \phi  \; = \;  \{ f  \circ \phi , g \circ \phi  \}  \; \; \forall f,g \in C^{\infty} (M) \}
\end{equation}
the hamiltonian flow in the Schr\"{o}dinger picture is made of symplectic diffeomorphisms and that the Liouville measure $  \mu_{Liouville} $  is $Symp (M , \omega )$-invariant.

\bigskip

Using the Moyal bracket instead of the Poisson bracket it is then natural to introduce the following:
\begin{definition}
\end{definition}
\emph{group of the quantum symplectic diffeomorphisms of $ ( M , \omega)$}:
\begin{equation}
    Symp_{\hbar} (M , \omega ) \; := \; \{ \phi \in Diff(M) \, : \,  \{  f ,  g \}_{\hbar} \circ \phi  \; = \;  \{ f  \circ \phi , g \circ \phi  \}_{\hbar}  \; \; \forall f,g \in C^{\infty} (M) \}
\end{equation}

The definition  \ref{def:Poisson brackets} naturally suggests to introduce the following:
\begin{definition}
\end{definition}
\emph{deformed symplectic form on $ ( M , \omega )$:}

the 2-form $ \omega_{\hbar} $ on M such that:
\begin{equation}
    \{ f , g \}_{\hbar} \; = \; \omega_{\hbar} ( X_{f} , X_{g} ) \; \; \forall f,g \in C^{\infty} (M)
\end{equation}
in terms of which it is natural to introduce the following:
\begin{definition}
\end{definition}
\emph{Moyal measure:}
\begin{equation}
    \mu_{Moyal} \; := \; \wedge_{i=1}^{\frac{dim M}{2}} \omega_{\hbar}
\end{equation}

Let us now observe that the quantum flow generated by the hamiltonian $ H \in C^{\infty} (M) $  may be seen as the commutative algebraic flow  $ F_{\hbar} (H)  := ( ( L^{\infty} ( M , \mu_{Moyal} ) , \omega_{\mu_{Moyal}} , \{  U_{t}(H)\}_{t \in \mathbb{R}} ) $ where the fact that the Moyal state $ \omega_{\mu_{Moyal}} $ is invariant under the quantum flow $  \{ U_{t}(H) \}_{t \in \mathbb{R}}$ is owed to the fact that the quantum flow in the Schr\"{o}dinger picture is made of quantum symplectic diffeomorphisms and that, by construction,  the Moyal measure $ \mu_{Moyal} $ is $Symp_{\hbar}(M , \omega )$-invariant.

\smallskip

We can finally define the quantum dynamical entropy as the Kolmogorov-Sinai entropy of the quantum flow:
\begin{definition}
\end{definition}
 \emph{quantum dynamical entropy of the quantum system with hamiltonian $ H \in C^{\infty}(M) $:}
 \begin{equation}
    h_{\hbar} (H) \; := \; h( F_{\hbar} (H) )
 \end{equation}

 Denoted by $ h(H):=  h( F (H) ) $ the Kolmogorov-Sinai entropy of the classical system with hamiltonian $  H \in C^{\infty}(M) $ we will say, according to the definition \ref{def:classical chaoticity}, that:
\begin{definition}
\end{definition}
 \emph{the classical system  with hamiltonian $ H \in C^{\infty}(M)$ is chaotic:}
\begin{equation}
    h(H) \; > \; 0
\end{equation}
while we will say that:
\begin{definition} \label{def:quantum chaos}
\end{definition}
\emph{the quantum system  with hamiltonian $ H \in C^{\infty}(M)$  is quantum-chaotic:}
\begin{equation}
    h_{\hbar} (H)  \; > \; 0
\end{equation}

\smallskip

\begin{remark}
\end{remark}
The definition \ref{def:quantum chaos} furnishes a new characterization of what Quantum Chaos is endowed with a clear information-theoretic meaning.

\end{document}